\documentclass[aps,twocolumn,showpacs,superscriptaddress]{revtex4}
\usepackage{graphicx,subfigure,amsmath,amssymb,multirow}

\begin{document}
\title{Novel mechanism for temperature-independent transitions in
flexible molecules: role of thermodynamic fluctuations}
\author{V. I. Teslenko}
\author{E. G. Petrov}
\affiliation{Bogolyubov Institute for Theoretical
Physics, Natl. Acad. Sci. of Ukraine,
14-b Metrologichna Str., 03680 Kyiv, Ukraine}
\author{A. Verkhratsky}
\affiliation{Faculty of Life Sciences, The University of Manchester, M 13 9PT Manchester,
UK}
\author{O. A. Krishtal}
\affiliation{Bogomoletz Institute of Physiology, Natl. Acad. Sci. of Ukraine, 4 Bogomoletz Str., 01601, Kyiv, Ukraine}

\begin{abstract}
 Novel physical mechanism is proposed for explanation of  temperature-independent transition reactions in molecular systems. The mechanism becomes effective in the case of conformation transitions between quasi-isoenergetic molecular states. It is shown that at room temperatures, stochastic broadening of molecular energy levels predominates the energy of low frequency vibrations accompanying the transition. This leads to a cancellation of temperature dependence in the stochastically averaged rate constants. As an example, physical interpretation of  temperature-independent onset of P2X$_3$ receptor desensitization in neuronal membranes is provided.

\end{abstract}

\pacs{02.50.Ey, 05.70.Ln, 87.10.Ed, 87.85.dm}

\maketitle

{\it Introduction}.-- Biomacromolecules (proteins and nucleic acids)
operate by maintaining the set of conformations and making transitions
between them with proper reaction rates which are, as a rule,
temperature controlled. Description of temperature-independent
effects requires dissection of multi-type reactions arising within
the complexity of molecular structures. In biology, one can generally
indicate three types of reactions. The first type is generally associated
with photoinduced oxidation-reduction processes
Basic mechanism of these reactions is an electron tunneling
between rigid redox centers like metal containing porphyrin rings
and quinones \cite{jort80}. The fact of tunneling is supported by the
absence of temperature dependence for the transfer rates in wide temperature
regions including the room temperatures \cite{warsh89,feh98}. The second
type of reactions is associated with  enzyme kinetics \cite{fern06} where
the rate constants are calculated using the concept of activated complex.
Motion of reactants along the reaction
coordinates is accompanied by overcoming the activation barriers
so that the rate constants experience exponential temperature dependence.
However, in the framework of Marcus's theory, a barrierless behavior becomes also possible if only reorganization energy and driving force of reaction are equal thus smoothing over temperature dependence.
Such situation is probably realized during the binding of NO molecule to the protoheme in the region of 200-290 K \cite{ion07}. The third type of
reactions is associated with conformation transformations. Some of
these reactions do not exhibit temperature dependence even in the range of room temperatures.  For instance, it has been demonstrated that the
closing rate for both the single-stranded DNA and RNA hairpin-loop
fluctuations  and the rate for the cyclic $\beta$ -hairpin peptide
folding are weakly  dependent on temperature in the interval
from $10^{\circ}$ to $60^{\circ}$C showing zero or even
slightly negative enthalpies \cite{wal01,kuz07}. Another example of
temperature independence in the gating
kinetics of membrane proteins has been recently  provided for
the desenstization onset of  P2X$_3$ purinoreceptors \cite{khm07}.

While physical mechanisms explaining both temperature-independent tunnel
electron transfer and barrierless  ligand binding are more or less clear
for two noted types of reactions, it is not the case for the third one.
But, we have to note here a recent paper of Pouthier \cite{pou09} devoted to the study of exciton diffusion in a lattice of H-bonded peptide units. As it follows from the paper,  a transition rate can become almost temperature independent in the strong anharmonic limit. Thus, the anharmonic nature of the phonons can be thought as one of possible mechanisms forming the  specific temperature behavior of transition rates.

In the present communication, we propose a novel physical mechanism explaining the existance of quasi-isoenergetic temperature-independent reactions in biosystems at room temperatures. It is assumed that for such type of reactions, the transitions between molecular states are accompanied by low-frequency molecular vibrations, while  molecular energies are alternated by thermodynamic fluctuations.

{\it Model and master equation}.-- Model is based on the
experimentally supported fact  that protein-structural motions
create electric field fluctuations accompanying the movements of
polar residues \cite{fay01}. Such structure fluctuations cause the
molecular energies to be stochastically time dependent and, thus,
the transitions between molecular states occur on the background of
random time-dependent energy shifts. Associating the molecule with
an open quantum system $S$, the respective molecular Hamiltonian can
be represented as $H(t)=H_0(t) +V$ where
$H_0(t)=\sum_n\,E_n(t)|n\rangle\langle n|$ is the main part of
Hamiltonian with $E_n(t)=E_n +\Delta E_n(t)$ being the state energy
varied owing to the stochastic addition $\Delta E_n(t)$. Transitions
between molecular states $|n\rangle$ and $|m\rangle$ are achieved by
nonadiabatic operator $V$. To analyze the experimental results one
have to know an evolution of observable occupancy $P_n(t)$ for each
molecular state involved into a given transition process. In our
case, this occupancy appears as an average over realizations of
random energy shifts. We denote such averaging via the symbol
$\langle\langle... \rangle\rangle$ so that
$P_n(t)=\langle\langle{\mathcal P}_n(t)\rangle\rangle$. In turn,
non-averaged occupancy ${\mathcal P}_n(t)$ can be found from exact
relation ${\mathcal P}_n(t)= \langle n|\rho (t)|n \rangle$ where
molecular density matrix $\rho (t)= tr_{B}\rho_{S+B}(t)$ is
determined as a trace (over the thermal bath states) on density
matrix $\rho_{S+B}(t)$ of an entire quantum system including a
thermal bath with respective bath Hamiltonian $H_B$ \cite{may04}.
Evolution of the $\rho_{S+B}(t)$ is governed by Liouville equation
$\dot{\rho}_{S+B}(t)=(-i/\hbar)[H_0(t)+V+H_B, \rho_{S+B}(t)]$. Using
this equation one can derive an exact master equation just for the
$\rho(t)$. For our purposes, however, it is quite enough to have a
master equation for only a diagonal part of molecular density
matrix, $\rho_d(t)=\sum_n\, \langle
n|\rho(t)|n\rangle|n\rangle\langle n|\equiv D\rho (t)$. In what
follow, we associate nuclear vibrations with a thermal bath. It
allows one to use a well known factorization
$\rho_{S+B}(t)=\rho(t)\rho_B$ where
$\rho_B=\exp{(-H_B/k_BT)}/tr_B\exp{(-H_B/k_BT)}$  is the bath
equilibrium density matrix. Using now a method of projection
operators \cite{zwa64} and generalizing this method for the case of
time-dependent Hamiltonian, we  achieve the following master
equation
%
\begin{eqnarray}
\dot{\rho}_d(t)=-\frac{1}{\hbar^2}\int_0^t d\tau\nonumber\\
\times tr_B\big\{D[V,U(t,t-\tau)[V,
\rho_d(t-\tau)\rho_B]U^+(t,t-\tau)]\big\} \label{me1}
\end{eqnarray}
where the evolution operator reads $U(t,t-\tau)=\hat{T}\exp{[-(i/\hbar)\int^t_{t-\tau}\,
d\tau'[H_0(\tau')+H_B+(1-D)V]}$ ($\hat{T}$ is the Dayson's chronological operator). Eq.
(\ref{me1}) is the basic coarse-grained stochastic equation which
can be applied for a rigorous description of transitions between
molecular states on the time scale $\Delta t\gg\tau_{vib}$ where  $\tau_{vib}$ is the characteristic time of establishment of a thermal equilibrium.

{\it Averaged transition rate constant}.-- In what follows we
concentrate our attention on the transition processes caused by a
weak coupling of molecular states to the vibrations. As it follows from a theory \cite{jort80,jort76}, if energy of a single phonon is able to cover a difference between energies of molecular states involved in the transfer process, the main contribution in transfer rate follows from single phonon-transitions while a role of multi-phonon processes becomes minor. Just such situation is assumed to be valid for low-energetic transitions in flexible molecules. The noted circumstances allow one to restrict a derivation to the Born approximation over nonadiabatic operator \cite{fong75}
$V=\sum_{n,m}\,(1-\delta_{n,m})\,\sum_{\lambda}\,\kappa^{\lambda}_{nm}\,(\beta_{\lambda}-
\beta^+_{\lambda})|n\rangle\langle m|$. Here, $\beta_{\lambda}$ (
$\beta^+_{\lambda})$ is the phonon annihilation (creation) operator
for  the $\lambda$th mode and
$\kappa^{\lambda}_{nm}=-\kappa^{\lambda}_{mn}$ is the coupling
constant.  Substituting the $V$ in master
equation (\ref{me1}) and taking the bath Hamiltonian in a convenient
form,
$H_B=\sum_{\lambda}\,\hbar\omega_{\lambda}(\beta^+_{\lambda}\beta_{\lambda}
+1/2)$ ($\omega_{\lambda}$ is the frequency of  $\lambda$th mode),
one achieves the following set of non-Markovian equations
%
\begin{eqnarray}
\dot{{\mathcal P}}_n(t)=-\frac{2{\rm Re}}{\hbar^2}\sum_m\,
\sum_{\lambda}\,|\kappa^{\lambda}_{nm}|^2\,\int_0^t\,d\tau \,R_{\lambda}(\tau)\nonumber\\
\times\big[e^{i\Omega_{nm}\tau}{\mathcal F}_{nm}(t,t-\tau) {\mathcal P}_n(t-\tau)\nonumber\\
-e^{i\Omega_{mn}\tau} {\mathcal F}_{mn}(t,t-\tau){\mathcal P}_m(t-\tau)\big]
\,.
\label{me2}
\end{eqnarray}
In this integro-diiferential equation, $\Omega_{nm}=(E_n-E_m)/\hbar$ is the transition frequency, ${\mathcal F}_{nm}(t,t-\tau)=\exp{\{(i/\hbar)\int_{t-\tau}^t\,
d\tau'[\Delta E_n(\tau')-\Delta E_m(\tau')]\}}$ is the stochastic
functional, and
$R_{\lambda}(\tau)=N(\omega_{\lambda})\exp{(i\omega_{\lambda}\tau)}+
[1+N(\omega_{\lambda})]\exp{(-i\omega_{\lambda}\tau)}$ is the regular single-phonon factor with $N(\omega_{\lambda})=[\exp{(\hbar\omega_{\lambda}/k_BT)-1}]^{-1}$
being the phonon distribution function.

To derive equations for the observable occupancies $P_n(t)$, one has to
average the stochastic equation (\ref{me2}). Here, we consider only the
case of fast variations for each random energy difference $\Delta
E_n(\tau) - \Delta
E_m(\tau)$ when the characteristic times of
stochastic and transition processes ($\tau_{stoch}$ and $\tau_{tr}$,
respectively) obey inequality $\tau_{stoch}\ll\tau_{tr}$. This
inequality allows us  to use a decoupling procedure $\langle\langle
{\mathcal F}_{nm}(t,t-\tau){\mathcal P}_n(t-\tau)\rangle\rangle=
\langle\langle {\mathcal F}_{nm}(t,t-\tau)\rangle\rangle
P_{n}(t-\tau)$. Note now that in the Born approximation, a
non-Markovian character of the transition process becomes not
important so that one can set $P_{n}(t-\tau)\approx P_{n}(t)$ (see
more details in ref. \cite{akh81}). Estimation of averages like
$\langle\langle {\mathcal F}_{nm}(t,t-\tau)\rangle\rangle
=F_{nm}(\tau)$ is given in numerous papers (see, e.g. refs.
\cite{bri74,sha78,pt91,goy97,egp98,gh}). The results indicate that
independently of a concrete form of quantities $F_{nm}(\tau)$, all
of them decay with own characteristic
times $\tau_{stoch}$. Taking this fact into account and using the
condition $\tau_{stoch}\ll\tau_{tr}$ one can shift the upper
limit in the integral from $t$ up to infinity. Thus, the averaging
over the fast random realizations transforms stochastic equation (\ref{me2})  into balance like kinetic
equation,
%
\begin{equation}
\dot{P}_n(t)=-\sum_m\,\big[K_{nm}P_n(t)-K_{mn}P_m(t)\big]
\label{me3}
\end{equation}
with the averaged transition rate constants
%
\begin{equation}
K_{nm}=\frac{2{\rm
Re}}{\hbar^2}\,\sum_{\lambda}\,|\kappa^{\lambda}_{nm}|^2\,
\int_{0}^{\infty}\, d\tau
e^{i\Omega_{nm}\tau}\,R_{\lambda}(\tau)F_{nm}(\tau)\,.\label{rate1}
\end{equation}

Now we consider an important random process which allows for an
exact calculation of the $F_{nm}(\tau)$. Let stochastically caused
alternation of  molecular energy differences be   characterized by
the only random quantity  $\alpha_{nm}(t)$ so that $\Delta
E_n(\tau)-\Delta E_m(\tau)= \hbar \alpha_{nm}(\tau)$. For the sake
of definiteness, let  random quantity fluctuate between two
equiprobable values $+\sigma$ and $-\sigma$ with mean frequency
$\nu$ (symmetric dichotomous process, DP) so that
$\alpha_{nm}(t)=\alpha(t)=\pm\sigma$.  Such random variation of
energy shifts yields $F_{nm}(\tau) = F(\tau)=\langle\langle
X(\tau)\rangle\rangle$ where
$X(\tau)=\exp{[i\int_{0}^{\tau}\,d\tau'\alpha(\tau')]}$ is the
stochastic functional.  Bearing in mind exact properties of the DP
so as $\alpha^2(\tau)=\sigma^2$, $\langle\langle
\alpha(\tau)\rangle\rangle=0$, and $\langle\langle
\alpha(\tau)\alpha(\tau_0)\rangle\rangle=\sigma^2\exp{(-\nu|\tau-\tau_0|)}$
and using the Shapiro-Loginov theorem \cite{sha78} in the form of
relation $d\langle\langle\alpha(\tau)X(\tau)\rangle\rangle/ d\tau
=i\nu d\langle\langle X(\tau)\rangle\rangle /d\tau
+i\sigma^2\langle\langle X(\tau)\rangle\rangle$, one obtains
$F(\tau)=(k_1e^{-k_2\tau}-k_2e^{-k_1\tau})/(k_1-k_2)$. Here,
quantities $k_{1,2}=\nu/2\pm\sqrt{(\nu/2)^2-\sigma^2}$ determine the
above noted characteristic stochastic times
$\tau_{stoch}^{(1,2)}=({\rm Re}\,k_{1,2})^{-1}$. Substituting the
$F_{nm}(\tau) =F(\tau)$ in Eq. (\ref{rate1}) and calculating the
respective integral one achieves the following expression for an
averaged transition rate constant
%
\begin{eqnarray}
K_{nm}=\frac{2\pi}{\hbar^2}\,
\sum_{\lambda}\,|\kappa^{\lambda}_{nm}|^2\Big [
N(\omega_{\lambda})\,L(\gamma,\nu;\Omega_{nm} +\omega_{\lambda})\nonumber\\
+(N(\omega_{\lambda})+1)\,L(\gamma,\nu;\Omega_{nm} -\omega_{\lambda})\Big]\,.
\label{rate2}
\end{eqnarray}
Here, the influence of stochastic field on the $n\rightarrow m$ transition is concentrated  in Lorentzian like function
$L(\gamma,\nu;\Omega)=\pi^{-1}\,
\gamma/[(\gamma -\Omega^2/\nu)^2
+\Omega^2]$
via parameters $\nu$ and $\gamma\equiv\sigma^2/\nu$.

{\it Temperature independence of rate constant}.-- Rate constant
(\ref{rate2}) includes a stochastic field in a non-perturbation
manner. This circumstance allows us to analyze various regimes of
transition processes in molecular systems dependently on value of
field parameters $\nu$ and $\gamma$.  Here, bearing in mind an
application of theory to explanation of conformational
quasi-isoenergetic transitions ($\Omega_{nm}\approx 0$) in flexible
biomolecules at room temperatures, we restrict ourself to the
analysis of transitions which are accompanied by low frequency
vibration modes so that $|\Omega_{nm}\pm\omega_{\lambda}|\ll\nu$.
This reduces function $L(\gamma,\nu;\Omega)$ to a standard
Lorentzian $L(\gamma,\Omega)\approx\pi^{-1}\gamma/(\gamma^2
+\Omega^2)$ where parameter $2\hbar\gamma$ can be treated as the
broadening of molecular energy levels. At $\gamma\rightarrow 0$,
i.e. in the case of extremely high frequency stochastic field
($\nu\rightarrow \infty$), Lorentzian is converted into the Dirac's
delta-function $\delta (\Omega)$ and, thus, rate constant
(\ref{rate2}) appears in a conventional form valid for Born
approximation over a nonadiabatic perturbation. In the case of
moderately high frequency stochastic field under consideration, an
another limit is realized where $\gamma\gg \Omega$. It yields
$L(\gamma,\Omega)=(\pi\gamma)^{-1}$. Note now that  room
temperatures correspond to energies of the order 0.025 eV, identical
to $6\cdot 10^{12}$s$^{-1}$ or 200 cm$^{-1}$. Therefore, if
transitions are accompanied by the vibrations of the order 60
cm$^{-1}$ and lower, one can set $N(\omega_{\lambda})\simeq
k_BT/\hbar\omega_{\lambda}$. This reduces averaged rate constant
(\ref{rate2}) to the form
%
\begin{equation}
K_{nm}=\frac{4k_BT}{\hbar^2\gamma}\,\sum_{\lambda}\,
\frac{|\kappa^{\lambda}_{nm}|^2}{\hbar\omega_{\lambda}}\,.
\label{rate3}
\end{equation}

Physical origin of stochastic parameter $\gamma$ is dictated by random shifts of molecular energy levels. We propose a phenomenological model where mean positive     and mean negative shifts ($+\sigma/2$ and $-\sigma/2$, respectively) result from thermodynamic fluctuations of an energy so that $(\sigma/2)^2=\overline{\delta E^2}/(2\pi\hbar)^2$. In line with a general theory of thermodynamic fluctuations in canonic ensembles,  the average of square of energy fluctuations is calculated as $ \overline{\delta E^2}=k_BT^2(\partial \overline{E}/\partial T)$ \cite{ hil56} with $\overline{E}$ being the mean energy of a system. At the same time, the average frequency of fluctuations is associated  with a linear frequency $\nu=\overline{E}/2\pi\hbar$ \cite{fern06}. Thus, a stochastic parameter can be estimated with  relation $\gamma=(2k_BT^2/\pi\hbar)(\partial \ln{\overline{E}}/\partial T)$. In classical limit under consideration, a mean energy per a separate degree of freedom, is \emph{exactly} equal to thermal energy $k_BT$. Therefore, independently of precise molecular system, a total mean  energy $\overline{E}$ is proportional to  a bath temperature. It yields $\gamma=2k_BT/\pi\hbar$. Consequently,
%
\begin{equation}
K_{nm}=\frac{2\pi}{\hbar}\,R_{nm}\,.
\label{rate4}
\end{equation}
In Eq. (\ref{rate4}), $R_{nm}=(2\pi\hbar)^{-1}\int_{-\infty}^{+\infty}\,(d\omega/\omega)\,
J_{nm}(\omega)$ is the temperature-independent factor which is expressed via spectral function $J_{nm}(\omega)=2\pi
\sum_{\lambda}\,|\kappa^{\lambda}_{nm}|^2\,
\delta(\omega_{\lambda}-\omega)$. The latter is widely used  in a spin-boson model for description of transport processes accompanied by vibration motions \cite{leg87}. In a given case, function $J_{nm}(\omega)$ includes an information on both a vibrational structure of flexible molecule and a character of nonadiabatic coupling to molecular vibrations.

\begin{figure}
\includegraphics[width=8cm]{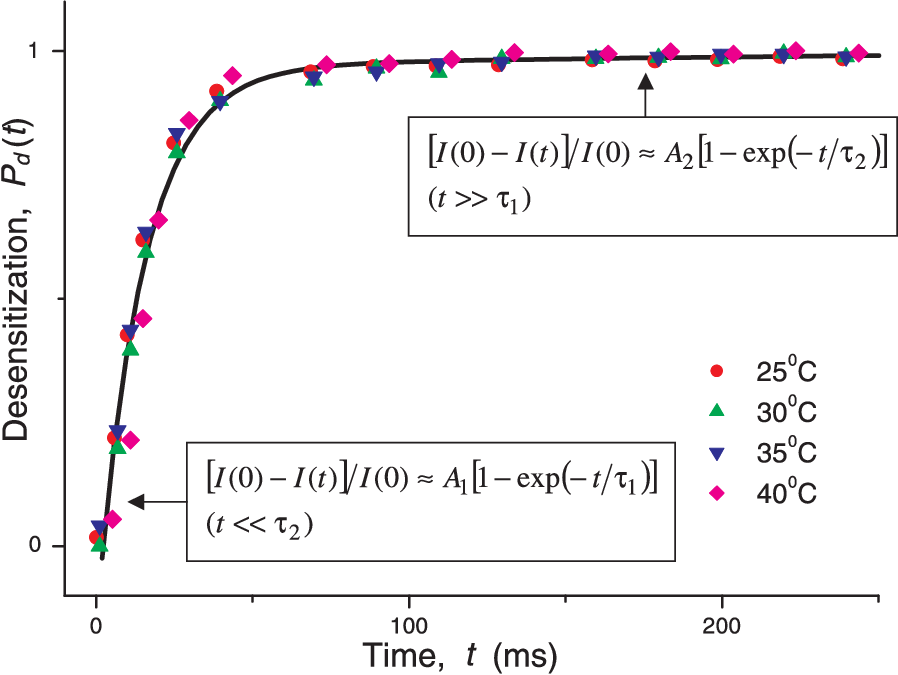}
\caption{Temperature independence of desensitization  onset of P2X$_3$ receptors measured at physiological temperatures $25^{\circ}$C-40$^{\circ}$C (selected from \protect\cite{khm07}) and its fit by Eq. (\ref{exp}). Insets indicate limiting cases in the model of two types of conducting
channels (details in the text).} \label{fig1}
\end{figure}

{\it Experimental evidence}.--Eq. (\ref{rate4}) contains a fundamentally important result indicating that the rates of quasi-isoenergetic transition processes in molecules can be temperature-independent even at room temperatures. As an example, we consider the desensitization onset  of P2X$_3$ receptors in neuronal membranes \cite{khm07}. For our knowledge, this is the first quantitative observation of temperature-independent gating in biological membranes.
Fig.  \ref{fig1}
manifests the decrease of ion current $I(t)$ through the  receptor channels. Desensitization probability of the channels, $P_d(t)=1-I(t)/I(0)$  is well described by a two-exponential kinetics \cite{khm07},
%
\begin{equation}
P_d(t)=1-A_1\,e^{-t/\tau_1}-A_2\,e^{-t/\tau_2}\,
\label{exp}
\end{equation}
where  $\tau_1\simeq 14.7$ms and $\tau_2\simeq 231$ms are the temperature-independent characteristic times of desensitization while $A_1\simeq 0.968$ and $A_2\simeq 0.032$ are the pre-exponential weights.
Since a particular molecular content of the P2X$_3$ receptor gates is yet unknown, we interpret the noted result in the framework of
simplest model supposing the formation of ion current as the sum of two partial currents. Let  $i_j$ be the current through a separate open channel of the  $j(=1,2)$th type. If   $N_j$ is the number  of respective ion channels, then a partial current  reads as $I_j(t)=N_j\,i_j\,P^{(j)}_{o}(t)$ where $P^{(j)}_{o}(t)$ is the probability for the channel to be in  the open state "$o$". Desensitization of each channel occurs independently of  one another manifesting the process of conformational transition from the open (conductive) state into the closed (nonconductive) state so that  $P^{(j)}_{o}(t)=\exp{(-t/\tau_j)}$. Introducing the quantities  $I(0)\equiv \sum_{j=1,2}\,N_j\,i_j\,$  and  $A_1=\xi/(1+\xi)$ and $A_2=1/(1+\xi)$ where
$\xi\equiv N_1\,i_1\,/N_2\,i_2\,$, one  expresses an ion current in the form $I(t)=I(0)\,P_o(t)$ where $P_o(t)=\sum_{j=1,2}\,A_j\,P^{(j)}_{o}(t)$ is the apparent (statistically averaged) probability of an ion channel to be in the open state. Bearing in mind that $P_{d}(t)=1-P_{o}(t)$, one arrives at the Eq. (\ref{exp}).

Physical explanation of temperature independency of  desensitization process can be given on the base of above proposed model of quasi-isoenergetic transitions in flexible molecules.
Denoting via $n=jo\alpha$ and
$m=jd\alpha'$ the conformational isoenergetic  substates participating in the open-close  transition, we introduce the integral occupancies of states $o$ and $d$ as $P_o^{(j)}(t)=\sum_{\alpha=1}^{\mu_{jo}}\,P_{jo\alpha}(t)$ and $P_d^{(j)}(t)=\sum_{\alpha'=1}^{\mu_{jd}}\,P_{jd\alpha'}(t)=1-P_o^{(j)}(t)$ where $\mu_{jo}$ and $\mu_{jd}$ are the number of respective degenerated substates for the $j$th type of channel.
Eq. (\ref{me3}) reads now as $\dot{P}_o^{(j)}(t)=-\big[K^{(j)}_{o\rightarrow d}P_o^{(j)}(t)-K^{(j)}_{d\rightarrow o}P_d^{(j)}(t)\big]$ where
$K^{(j)}_{o\rightarrow d}=K_j/\mu_{jo}$ and $K^{(j)}_{d\rightarrow o}=K_j/\mu_{jd}$ are the integral rate constants with $K_j\equiv (2\pi/\hbar)\,\sum_{\alpha=1}^{\mu_{jo}}\,\sum_{\alpha'=1}^{\mu_{jd}}\,
R_{jo\alpha\,jd\alpha'}$.
Since the disordering increases the degeneracy of molecular state, then
$\mu_{jd}\gg \mu_{jo}$. This reduces the kinetics of isoenergetic transitions to the nonrecurrent single-exponential kinetics
with the characteristic time $\tau_j\simeq (K^{(j)}_{o\rightarrow d})^{-1}$, for the $j$th type of channel. Thus, physically, the receptor desensitization appears as a temperature-independent open$\rightarrow$close transition process between quasi-isoenergetic molecular conformations of ion channel.

{\it Conclusion}.-In this communication,  a novel physical mechanism is proposed to explain the formation of temperature-independent transition processes in molecular systems. In contrast with well established mechanism describing quantum phononless site-to-site particle tunneling at low temperatures or  tunneling at room temperatures that is accompanied by an emission of  high frequency phonons (when $\omega_{\lambda}\gg  k_BT/\hbar$)\cite{jort80}, the proposed mechanism works in a  classic region of temperatures with participation of low frequency phonons (when $\omega_{\lambda}\ll k_BT/\hbar$). The lack of temperature dependence in  rate constant (\ref{rate2}) occurs owing to the thermodynamical stochastic variation of energy levels participating in the transitions. Frequency of this variation, $\nu$ is assumed to be fast in comparison to the transition frequencies $\Omega_{nm}\sim \omega_{\lambda}$. This reduces
an averaged rate constant (\ref{rate2}) to a  more simple form (\ref{rate3}) which, in turn, is simplified to the temperature-independent form (\ref{rate4}). Of course, a given mechanism is not unique and can work independently of or in parallel with other possible mechanisms, especially with anharmonic mechanism noted in ref. \cite{pou09}. But, for understanding the physics of quasi-isoenergetic transitions in  biological macromolecules, the reference of fast molecular motions to the motions responsible for the creation of high frequency stochastic fields, is assumed to be quite fruitfull from a  semi-phenomenological physical point of view.

{\it Acknowledgments}--
This work was supported in part by project of Fundamental Researches of NASU to VT and EG, and INTAS  grant to AV and OK.

\end{document}